%% file: ms.tex
\documentclass[conference]{IEEEtran}
\usepackage{fancybox}
\usepackage[ruled,vlined]{algorithm2e}
\usepackage{graphicx}
\usepackage{paralist}
\usepackage{tabularx}
\usepackage[hyphens]{url}
\usepackage{balance}
\usepackage{multirow}
\usepackage{multicol}
\usepackage{pbox}
\usepackage{mathtools}
\usepackage{cite}
\usepackage{color, colortbl}

\usepackage{tikz}
\usetikzlibrary{arrows}

\usepackage[TABBOTCAP]{subfigure}
\usepackage{url}
\usepackage{balance}
\usepackage{amsmath}
\usepackage{verbatim}
\usepackage{float}
\usepackage{colortbl}
\usepackage{amssymb}
\usepackage{ifthen}
\usepackage{pifont}
\include{macro}

\begin{document}
\newenvironment{myquote}{\list{}{\vspace{0.03in}\leftmargin=0.12in\rightmargin=0.12in}\item[]}{\endlist}

\title{How do Developers Test Android Applications?}

\author{\IEEEauthorblockN{Anonymous Authors}}

\author{\IEEEauthorblockN{Mario Linares-V\'asquez$^1$, Carlos Bernal-C\'ardenas$^2$, Kevin Moran$^2$,  and Denys Poshyvanyk$^2$}
\IEEEauthorblockA{
$^1$Universidad de los Andes, Bogot\'a, Colombia\\
$^2$College of William \& Mary, Williamsburg, VA, USA\\
m.linaresv@uniandes.edu.co,  \{cebernal, kpmoran, denys\}@cs.wm.edu}}

\maketitle

\begin{abstract}
Enabling fully automated testing of mobile applications has recently become an important topic of study for both researchers and practitioners. A plethora of tools and approaches have been proposed to aid mobile developers both by augmenting manual testing practices and by automating various parts of the testing process. However, current approaches for automated testing fall short in convincing developers about their benefits, leading to a majority of mobile testing being performed manually. With the goal of helping researchers and practitioners -- who design approaches supporting mobile testing -- to understand developer's needs,  we analyzed survey responses from 102 open source contributors to Android projects about their practices when performing testing. The survey focused on questions regarding practices and preferences of developers/testers in-the-wild for (i) designing and generating test cases, (ii) automated testing practices, and (iii) perceptions of quality metrics such as code coverage for determining test quality. Analyzing the information gleaned from this survey, we compile a body of knowledge to help guide researchers and professionals toward tailoring new automated testing approaches to the need of a diverse set of open source developers.
\end{abstract}

\IEEEpeerreviewmaketitle

\vspace{-0.2cm}
\section{Introduction}
\label{sec:intro}
\input{intro}

\section{Design of the Empirical Study}
\label{sec:study}
\input{study}

\section{Results and Discussion}
\label{sec:approach-uc}
\input{approach-uc}

\section{Threats to Validity}
\label{sec:threats}
\input{threats}

\section{Related Work}
\label{sec:related}
\input{related}

\section{Conclusion and Learned Lessons}
\label{sec:concl}
\input{concl}

\vspace{-0.2cm}

\balance
\bibliographystyle{abbrv}
\bibliography{ms}

\end{document}

%% file: intro.tex
	Mobile devices have quickly become the most accessible and popular computing devices in the world \cite{smartphone-popularity} due to their affordability and intuitive, touch-based user interfaces. The ubiquity of smartphones and tablets has led to sustained developer interest in creating ``apps" and releasing them on increasingly competitive marketplaces such as Apple's App Store \cite{AppleStore} or Google Play \cite{google-play}.  Due to their highly gesture-driven nature, GUI-based testing of mobile apps is paramount to ensuring proper functionality, performance, and an intelligent user experience.  However, GUI-based testing activities are typically costly, and in the context of mobile apps, often performed manually \cite{Mona:ESEM13,Kochhar:ICST15}. Given the additional constraints on the mobile application development process including pressure for frequent releases \cite{Hu:EuroSys14,Jones:2014}, rapid platform evolution and API instability \cite{Bavota:TSE15,Linares-Vasquez:FSE13}, and parallel development across different platforms \cite{Mona:ESEM13,Ali:MobileSoft17} it can be difficult for developers to budget time for effective testing.  Thus, the challenge of automating mobile app testing has captured the interest of the software engineering and systems research communities, and has led to the development different types of automated techniques that assist in various testing tasks.

	Research-oriented tools aimed at improving mobile testing span a diverse range, from record \& replay approaches \cite{Gomez:ICSE13,Hu:OOPSLA2015}, to bug reporting aids \cite{Moran:FSE15,Moran:ICSE16}, to automated input generation techniques \cite{android-monkey,Machiry:FSE13,Sasnauskas:WODA14,Ravindranath:Mobisys2014,Amalfitano:ASE12,Anand:FSE12,Azim:OOPSLA13,Moran:ICST16,google-robo-test,Amalfitano:IEEE14,Azim:OOPSLA13,Choi:OOPSLA13,Zaeem:ICST2014,Yang:FASE13,Linares:MSR15,Zhang:ICSE17,Jensen:ISSTA13,Mirzaei:ISSRE15,Mahmood:FSE14,Mao:ISSTA16}. Perhaps the most interesting and valuable of these techniques from a developer's or tester's  perspective are the automated input generation (AIG) techniques.  The high-level goal of such techniques is relatively simple: given a mobile application under test (AUT), generate a series of program \textit{inputs} according to a pre-defined \textit{testing goal}.  For the vast majority of these techniques, the generated \textit{inputs} are simulated touch events on the screen of a device, and the \textit{testing goal} is typically either achieving the highest possible code-coverage or uncovering the highest number of faults (e.g., crashes).   

	However, despite the large amount of research effort dedicated to building AIG techniques and other automated approaches, recent studies seem to indicate that these approaches are typically not used in practice \cite{Kochhar:ICST15}.  Choudhary et. al offer a set of potential reasons for this lack of adoption as part of an experience report analyzing several AIG tools \cite{Shauvik:2015}, and among the reasons reported are (i) lack of reproducible test cases, (ii) side effects across different testing runs, and (iii) lack of debugging support. While this study offers some insight, researchers and practitioners who aim to build these tools with the intention of them gaining adoption and positively impacting developers do not have a clear understanding of the testing needs and preferences of real developers. 

	This lack of guiding direction for this particular topic of mobile software engineering research is somewhat troubling given the highly practical impact that such tools could have on daily development and testing workflows.  Conversely, it is unsurprising that many tools have failed to make an impact without taking into account developer preferences, as \textit{``Automation applied to an inefficient operation will magnify the inefficiency"}\footnote{Bill Gates, co-founder of Microsoft, in reference to automation in business settings}.  If the \textit{operation} or \textit{goals} of automation techniques do not match developer needs, preferences, and expectations (and are thus \textit{inefficient}), there is little chance that these will have a meaningful impact.  Therefore, there is a very clear demand to align the goals of research on automated testing techniques with the needs of developers in order to allow for practical impact.

	In this paper, we aim to bridge this gap through a survey, that at its core, aims to examine the testing preferences of open source developers with the intended purpose of providing actionable information to researchers and practitioners working on approaches to automate different aspects of mobile testing.    

In summary, this paper makes the following noteworthy contributions:

\begin{itemize}

\item To the best of our knowledge, this is the first paper aimed at analyzing mobile testing preferences of real open source developers with a focus on (i) typical preferences when designing test cases for mobile apps, (ii) preferred characteristics for automatically generated test cases, and (iii) preferred effectiveness metrics.

\item This study complements previous work that has identified and speculated upon potential reasons for lack of adoption of automated mobile testing approaches by collecting information from open source developers and providing a set of learned lessons to guide future research.

\item Our general findings indicate that developers (i) rely heavily on usage models of their applications when designing test cases, (ii) prefer high-level, expressive automatically generated test cases organized around use-cases, and (iii) prefer manual testing over automation due to factors including test case representation and issues with reproducibility (iv) do not hold the perception that code coverage is an important measure of test case quality, as indicated by a large portion ($\approx$ 64\%) of study participants, instead citing other measures of quality such as feature coverage or fault-detection as more useful.
 
\end{itemize}

%% file: study.tex
The main \emph{goal} of this study is to identify and analyze practices and preferences of mobile developers (MDs) toward testing related activities. To this end, we explored MDs (i) practices in documenting requirements and designing test cases, (ii) preferences toward features of automated testing approaches, (iii) use of existing automated tools, and (iv) preferences for testing-related quality measures.  The study is intended to benefit the \textit{perspective} of researchers and practitioners interested in designing approaches and tools for automated testing of mobile apps. 

While common wisdom and best practices suggest that test cases should be derived from requirements artifacts,  to the best of our knowledge, previous studies focused on the challenges and tools used for testing but without analyzing the details of the strategies used by MDs for designing manual test cases or exploring preferences for automatically generated test cases \cite{Mona:ESEM13, Kochhar:ICST15}\footnote{Kochhar \emph{et al} \cite{Kochhar:ICST15} investigated also with a survey with 83 open source Android developers the tools they use and challenges they face while testing Android apps. It is worth noting that our survey also includes a question concerning the tools used for automated testing (See \textbf{SQ8} in Table~\ref{tab:survey1-questions}). Kochhar \emph{et al} \cite{Kochhar:ICST15} list a reduced set of tools (i.e., 10), however, we complement their findings with a list of 55 tools used by our surveyed participants. We also complement their findings with a specific question designed to understand experiences and issues of mobile developers when using random testing tools (See \textbf{SQ9} in Table~\ref{tab:survey1-questions})}. These aspects are important, as learning the preferences of developers' manual testing practices can inform automated techniques to best meet these needs.

Consequently, we aim to fill this ``gap'' in recent work by surveying contributors of  open source Android apps hosted on GitHub. We are most interested in understanding testing practices of mobile developers from the viewpoint of test case design, preferred testing strategies, reasons for the prevalence of manual testing (as suggested by previous work \cite{Mona:ESEM13, Kochhar:ICST15}), and preferred information/features of ideal automatically generated test cases. We also wanted to survey MDs about the usage of widely used techniques in the research community such as random testing and coverage analysis. 

\subsubsection{Research Questions} In particular, we aimed at answering the following research questions (RQs):

\begin{myquote}
\noindent \textbf{RQ$_1$}: \emph{What are the strategies used by MDs to design test cases?}\\
\noindent \textbf{RQ$_2$}: \emph{What are the MDs' preferences for automatically generated test cases?}\\
\noindent \textbf{RQ$_3$}: \emph{What tools are used by MDs for automated testing?}\\
\noindent \textbf{RQ$_4$}: \emph{Do MDs consider code coverage as a useful metric for evaluating test cases effectiveness?}\\
\end{myquote}

\begin{table}[t]
\caption{Survey Questions for our Empirical Study}
\label{tab:survey1-questions}
\vspace{-0.5cm}
\begin{center}
\begin{tabular}{l |m{7cm}}
\hline
\textbf{Id} &\textbf{Question (Type)}\\
\hline
 \rowcolor[gray]{.9}SQ$_1$&What type of documentation do you use for specifying the requirements in your apps? (Multiple choice)\\

SQ$_2$&How do you usually distribute your testing time among these different activities (e.g., manual testing 20\%, Junit testing 50\%, cloud testing 30\%)? (Open question)\\ 

\rowcolor[gray]{.9}SQ$_3$&Please provide rationale for your answers to SQ3. (Open question)\\ 

SQ$_4$& How do you design the test cases for your apps? (Open question)\\ 

\rowcolor[gray]{.9}SQ$_5$& What is the target of the test cases you design for your apps? (Multiple choice)\\ 

SQ$_6$&If you are using (or intend to use) tools for automatic generation of test cases, what format for the test cases do you prefer? (Single choice)\\ 

\rowcolor[gray]{.9}SQ$_7$&Assuming you have test cases in natural language, what type of information would you like to have in them? (Multiple choice)\\ 

SQ$_8$&What tools do you use for automated testing? (Open question)\\ 

\rowcolor[gray]{.9} SQ$_9$&What are your experiences with random testing tools such as Android Monkey? Are random testing tools useful for your needs? Did you experience any issues with the sequences of events generated by a random testing tool like Android Monkey? (Open question)\\ 

SQ$_{10}$&Do you use code coverage as a metric for measuring the quality of your test cases? Why? (If you answer is No, please describe how else you measure or ensure the quality of your test cases) (Open question)\\ 

\hline
\end{tabular}
\end{center}

\end{table}

\subsubsection{Data Collection} Table \ref{tab:survey1-questions} lists the questions in the survey.  SQ$_1$-SQ$_5$ were used to answer \textbf{RQ$_1$}; SQ$_6$ and SQ$_7$ aim at answering \textbf{RQ$_2$}; SQ$_8$ and SQ$_9$ were designed to answer \textbf{RQ$_3$} ; and SQ$_{10}$ served to answer \textbf{RQ$_4$}. We also collected demographic background information to filter participants with short or over claimed experience in Android development, and to measure the diversity of our sample.   

The survey was hosted online on the Qualtrics platform~\cite{Qualtrics}, and the participants were contacted via email. To select the potential participants, we followed the same procedure from previous work ~\cite{Linares-Vasquez:ICSME15}  --- that also surveyed open source developers --- to extract contributors' emails from GitHub. After the extraction and filtering, we emailed the survey to 10,000 email addresses from which we got 485 survey responses. We discarded 5 responses in which the participant reported 0 years of experience in Android programming, 3 with invalid answers, and 370 unfinished surveys. In the end, we obtained 102 valid responses. 

\begin{figure}[t]
\begin{center}
\includegraphics[width=\linewidth]{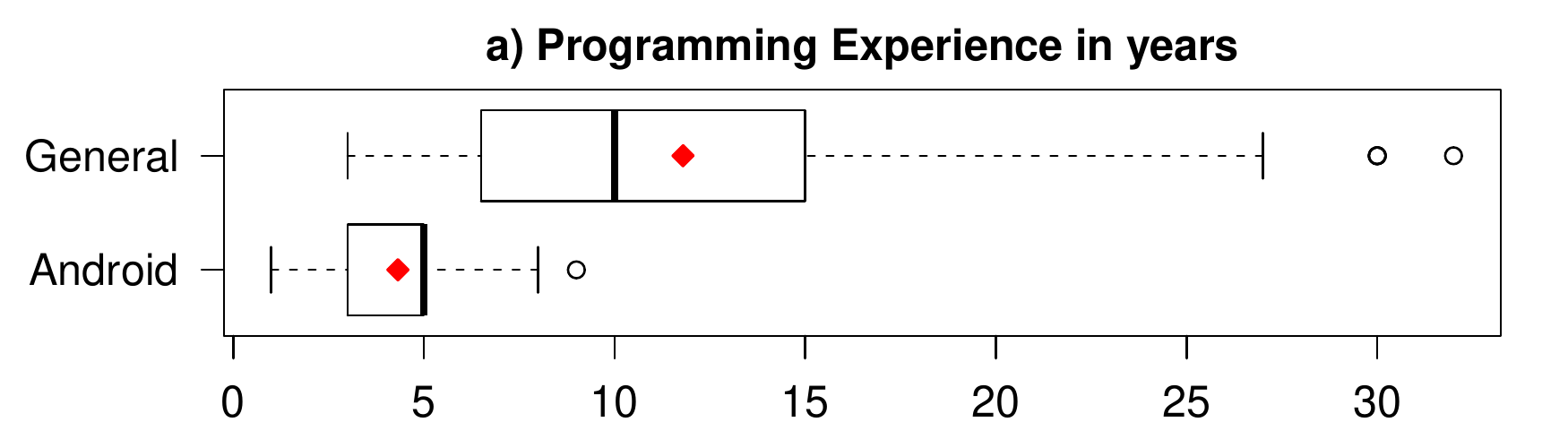}
\includegraphics[width=\linewidth]{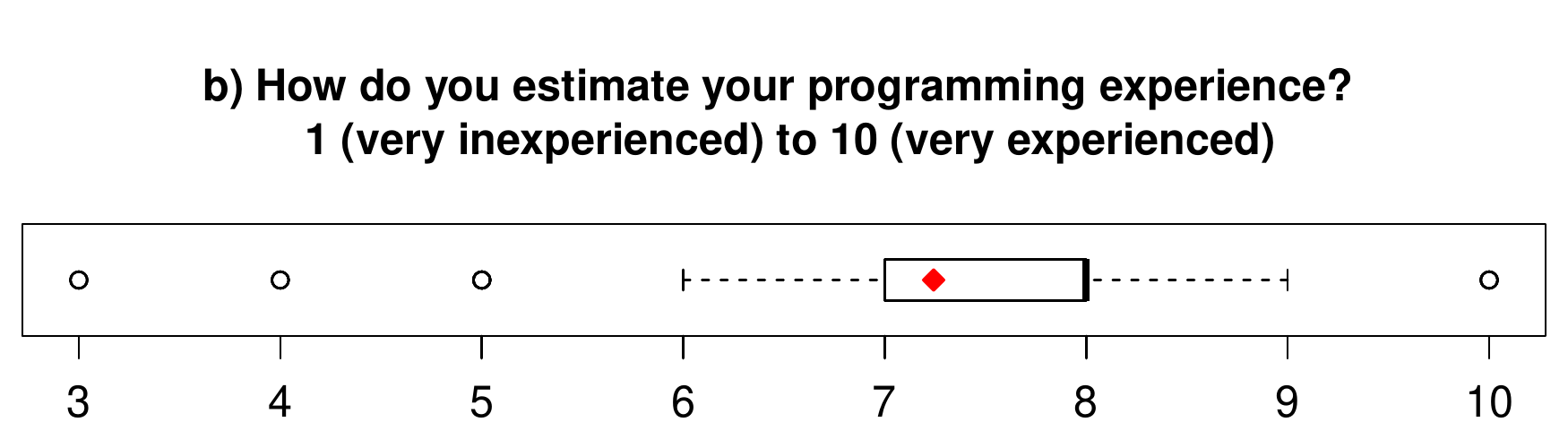}
\includegraphics[width=\linewidth]{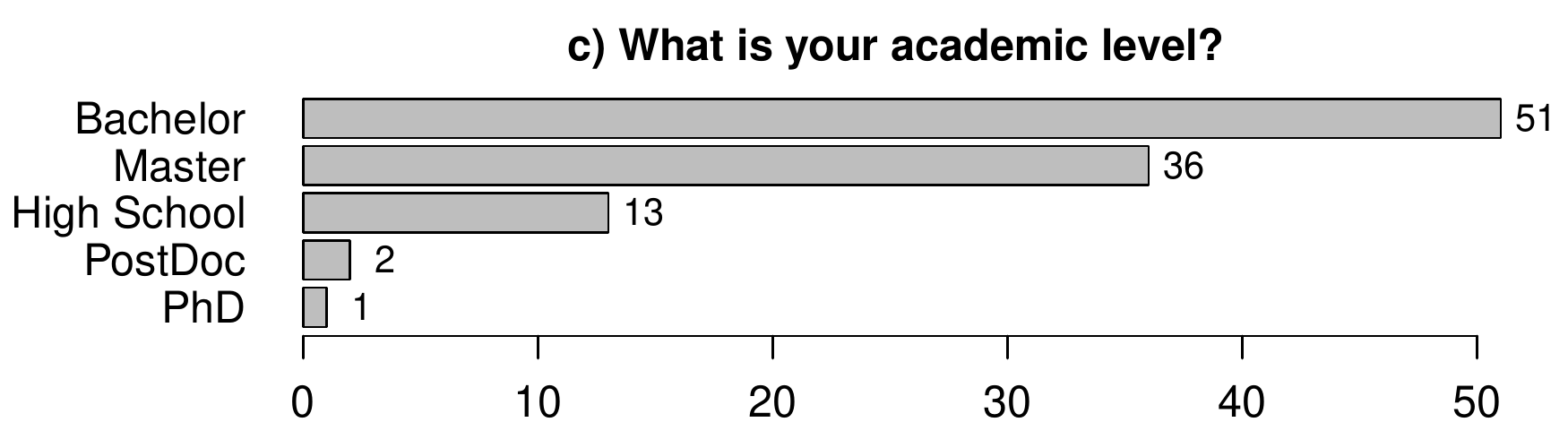}
\includegraphics[width=\linewidth]{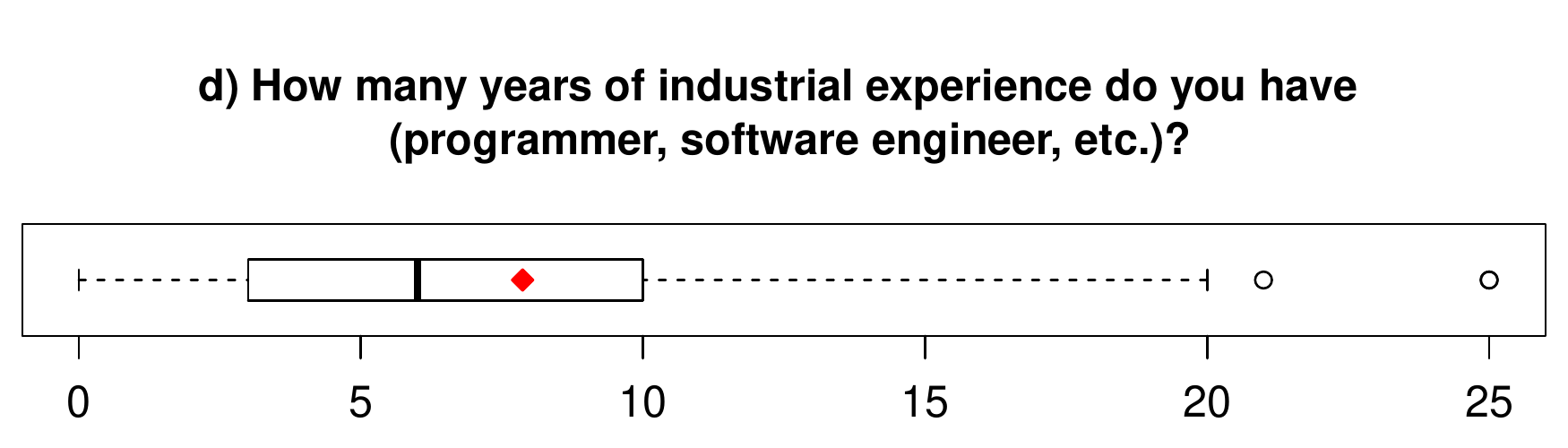}
\vspace{-0.8cm}
\caption{Results of demographic questions for the 102 survey participants}
\label{fig:demo-survey}
\end{center}
\end{figure}

The demographics of the participants are depicted in Figure~\ref{fig:demo-survey}. Our sample (102) is comparable to final numbers of mobile developers (with industrial experience) surveyed/interviewed in previous studies investigating other software engineering phenomenon:  3 in~\cite{Linares:FSE15}, 9 in~\cite{Miranda:CMSES14}, 45 in~\cite{Bavota:TSE15}, 83 in~\cite{Kochhar:ICST15}, 200 (188 developers + 12 experts) in~\cite{Mona:ESEM13}, and 485 in~\cite{Linares-Vasquez:ICSME15}. In addition, the claimed programming experience is diverse for the three cases: general programming, Android programming, and industrial experience.

The answers to multiple/single-choice questions were analyzed using descriptive statistics. In the case of open questions, we categorized the answers manually following a grounded theory-based approach~\cite{CorbinStrauss}. Three of the authors went through all of the free-text answers and performed one round of open coding by independently creating categories for the answers. After the round of open-coding, the codes were standardized. In the cases of non-agreement between the three coders, corresponding answers were marked as ``Unclear''.

%% file: approach-uc.tex
	In this section we report the responses by the participants and provide answers to the aforementioned research questions. We describe the results using descriptive statistics and through summaries and  discussion of examples of free-text answers.

\subsection{RQ$_1$: What are the strategies used by MDs to design test cases?}

\begin{figure}[t]
\begin{center}
\includegraphics[width=\linewidth]{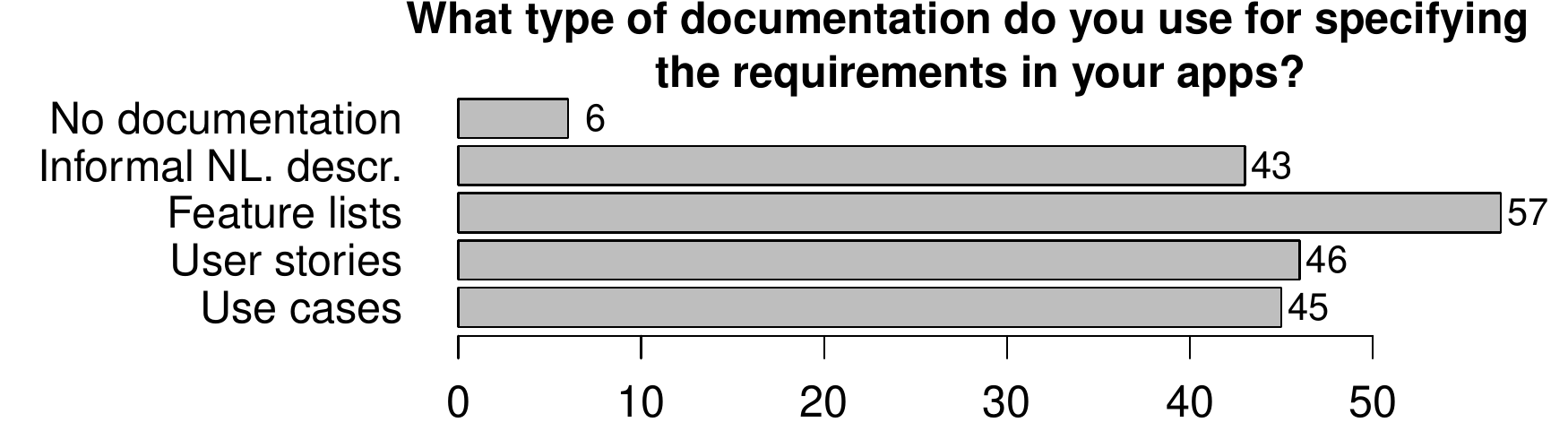}
\vspace{-0.8cm}
\caption{Artifacts used by Android developers to document apps requirements. The bar plots show the number of times each artifact was selected by the participants.}
\label{fig:documentation-survey}
\end{center}
\end{figure}

	\textbf{Artifacts for documenting app requirements}.  Android developers use a diverse set of artifacts to document requirements, including artifacts from disciplined and agile methods.  Fig. \ref{fig:documentation-survey} depicts the answers for each of the options in our \textbf{SQ1} (note that this was a multiple-choice question).  Although there is no large tendency towards a preferred artifact, feature lists are the top-used artifact. Surprisingly, only six participants do not document requirements, and the usage of the other options (i.e., use cases, user stories, and informal natural language descriptions) is balanced across the participants. When analyzing the most popular answers (including combination of choices as a whole answer provided by participants), there is no clear preference; however, we found that the most popular responses reporting the usage of only a single type of artifact as the documentation practice are distributed as follows:  user stories (14  participants), feature lists (12), informal natural language descriptions (10),  use cases (9). Additionally some participants selected the combination of all 4 artifacts (8) as a single answer. 

\begin{figure}[h]
\begin{center}
\includegraphics[width=\linewidth]{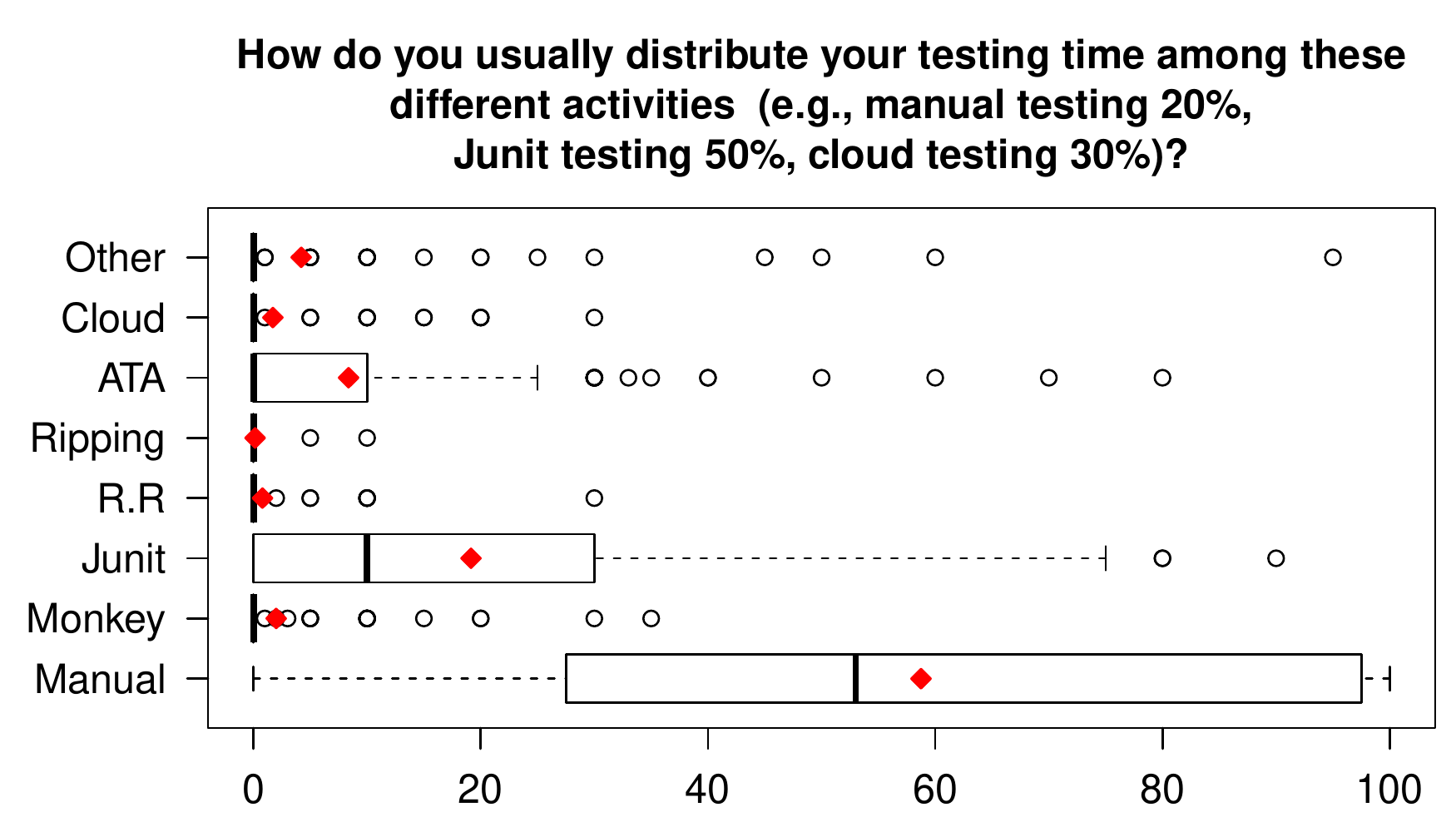}
\vspace{-0.8cm}
\caption{Distribution of testing activities reported by our participants.  *ATA means Automated Testing APIs, and R.R means Record \& Replay.}
\label{fig:test-effort-distr}
\end{center}
\end{figure}

	\textbf{Distribution of testing efforts}. Previous studies have reported that manual testing is preferred over automated approaches \cite{Mona:ESEM13, Kochhar:ICST15}. In the survey, we asked participants about how they distribute their testing effort and time across different testing activities (\textbf{SQ2}). In particular we asked about the following activities: manual testing, random testing using Monkey, JUnit testing, Record \& Replay-based (R\&R) testing, GUI ripping-based tools, automated testing with automated testing APIs (ATA), cloud testing services, and others. The answers provided by the participants are depicted with boxplots in Fig.~\ref{fig:test-effort-distr}.
\vspace{-0.7cm}
\vspace{-0.4cm}

As expected, manual testing is the preferred testing activity with an average of 58.18\% of the testing efforts dedicated by the participants, an interquartile range (IQR) of $[25\%,97.25\%]$ and a maximum of 100\%; 96 out of 102 participants reported more than 5\% of testing effort devoted to manual testing with 35 of them reporting more than 90\% of dedication to manual testing. The second most popular activity in which the developers dedicate their testing efforts is JUnit-based testing with and average dedication of 18.96\%, an IQR of $[0\%,30\%]$ and a maximum of 90\%;  58 participants reported more than 5\% of testing effort devoted to unit testing.  The third most popular testing activity is automated testing with APIs such as Expresso and Robotium, with an average dedication of 8.29\%, an IQR of $[0\%,10\%]$ and a maximum of 80\%;  34 participants reported more than 5\% of effort devoted to manual testing. 

Fuzz/Random testing with Android Monkey is widely used as the baseline for comparing new automated testing tools proposed by the research community~\cite{Machiry:FSE13,Linares:MSR15,Shauvik:2015,Moran:ICST16}. However, responses from our participants suggest that fuzz/random testing is not widely used in-the-wild. Only 11 participants reported more than 10\% of testing effort with Android Monkey with an average of 16.36\% (for those 11 participants) and a maximum of 35\%. One automated strategy that is widely used in the research community and serves as the foundation for multiple methods of AIG is GUI ripping~\cite{Amalfitano:ASE12,Azim:OOPSLA13,Amalfitano:IEEE14,Nguyen:TSE14}; however, it seems that developers in-the-wild are not aware of such tools or do not find them useful. Only two participants reported the usage of automated ripping tools  with 5\% and 10\% of their testing efforts. Concerning  the case of Record \& Replay, five participants reported this technique with more than 5\% of testing effort and a maximum of 30\%.

	Despite of the availability of services such as Xamarin test cloud~\cite{XamarinTC}, Saucelabs~\cite{saucelabs}, and Perfecto~\cite{Perfecto}, only 14 participants mentioned the usage of cloud services for testing with an average effort of 12.57\% (for those 14 participants) and a maximum of 30\%. 

	Finally, under the ``Others'' option, we got 14 answers in which the participants claimed more than 10\% of testing effort with an average of 29.29\% and maximum of 95\%. The participants further explained in the answer to \textbf{SQ3} that this is mostly because they use customized tools or strategies, instrumentation-based testing, or beta testing with users.

	Regarding the rationale provided by the participants for their choices, the preference for manual testing is supported by several reasons such as (i) changing requirements, (ii) lack of time for testing and process decisions, (iii) size of the apps, (iv) lack of knowledge of automated tools and techniques, (v) usability and learning curve of available tools for automated testing, and (vi) the cost of maintaining automated testing artifacts.  For instance, the following rationale provided by some participants illustrate their preference for manual testing as a consequence of changing requirements:

\begin{myquote}\emph{
``our app changes very frequently and we can't afford unit testing and automated testing''
}\end{myquote}

\begin{myquote}\emph{
``it's hard to right the useful test case for current project, because the requirement is changed very often,  and the schedule is very tiny. So we still prefer hire some tester to do manually testing.  In the other hand, the android test framework is not good enough yet, I tried study roboeletric, it's a little bit hard to understand.''
}\end{myquote}

\begin{myquote}\emph{
``A lot of what I do is related to how the app looks and feels. Therefore a lot of my testing is done manually. When I have complexity in my classes I use junit. Sometimes I use the Instrumentation testing classes from Android, but not much as manually testing feels faster. Also the requirements tend to change a lot so manually testing unfortunately is the best option for me.''
}\end{myquote}


The survey participants also justified the usage of manual testing because of time-related issues and project management decisions. For example:


\begin{myquote}\emph{
``Usually  customers doesn't provide enough time for development of automated tests.''
}\end{myquote}

\begin{myquote}\emph{
``Too much time required in configuring the components for automated testing.  Partly because I was working in a consultancy firm so there was no incentive to spend time on automated testing (not chargable)''
}\end{myquote}

\begin{myquote}\emph{
``I do not agree about this method but this is management decision. I repeatedly expressed by disconformity with this methodology.''
}\end{myquote}

\begin{myquote}\emph{
``Although I strongly disagree with this: the institution I work at does not provide the atmosphere to make testing a vital part of our development.''
}\end{myquote}

\begin{myquote}\emph{
``My previous work didn't have any testing requirements for the apps, and writing tests takes time that the budget didn't account for. ''
}\end{myquote}


	Cost, in terms of money and time, for creating and maintaining automated testing artifacts is also another reason for preferring manual testing. This case is illustrated by the following examples of rationale provided by the survey participants:


\begin{myquote}\emph{
``I'm faster by testing the app and all it's possibilities on the device itself, instead of writing separate Test Cases.''
}\end{myquote}

\begin{myquote}\emph{
``Quickly testing the product is important and writing automated tests can't be done quick enough. So a majority of my time is spent making sure the product works manually, then spending time automating what I can. jUnit is wrote by developers, so I've added a couple of tests, but not much. Have just played around with Android Monkey. Our product uses the Cloud, but another person takes care of the majority of the testing.''
}\end{myquote}

\begin{myquote}\emph{
``We prioritize feature work over automated testing. Automated tests have done little to prevent bugs, but incur significant overhead when creating new features or refactoring old.''
}\end{myquote}

\begin{myquote}\emph{
``For Android traditionally it has been very difficult to write tests for. Also UI tests can be brittle and take significant effort to maintain. In an environment where development resources are constrained and features / bug fixes works takes priority, it is very difficult to have sufficient tests. Manual testing with a dedicated QA team is more practical to maintain.''
}\end{myquote}

\begin{myquote}\emph{
``I'm skeptical of UI testing because I've found that the tests are fragile, require a lot of maintenance, and are generally more work than worth.''
}\end{myquote}


	The survey participants also claimed a general lack of knowledge of existing automated testing techniques, along with difficulties related to the usability of the tools as factors for not performing automated testing:

\begin{myquote}\emph{
``We don't really know how to use other tools.''
}\end{myquote}

\begin{myquote}\emph{
``I did not know about Android Monkey prior to this survey.  I will try it out.''
}\end{myquote}

\begin{myquote}\emph{
``I have not found any easy-to-use testing solutions for Android''
}\end{myquote}

Finally, the size and maturity of the apps is also a factor that influences the preference for manual testing:


\begin{myquote}\emph{
``I never got involved in an Android project big enough to require unit testing, it wasn't worth investing in that.''
}\end{myquote}

\begin{myquote}\emph{
``I mostly do manual testing due to the limited size of my apps. I sometimes use a custom replay system (built into the app) to duplicate bugs after I come across them. This method is usually combined with manual testing (printing debug information to the log) to pinpoint the cause.''
}\end{myquote}

\begin{myquote}\emph{
``I'm mostly just building toys or research prototypes, never built Android apps professionally. So I test pretty informally (and poorly) because I just want to build a thing quick and don't care if it's robust.''
}\end{myquote}


\textbf{Test case design strategies}. After the open coding for the responses to \textbf{SQ4}, 34 answers were not considered because (i) the participants explicitly mentioned they do not perform testing or do not design test cases, and (ii) for some answers we were not able to understand/codify the textual answer. From the valid/accepted answers, the top strategy reported by the participants to design test cases is follow the usage model of the app as a guide (30 answers). The next in the list is designing unit tests for individual components/methods (10 answers), followed by negative testing and edge cases (9 answers), and testing expected outputs (6 answers). Bugs, changes in the last version, and regression account for  9 answers. Three participants claimed they follow the Behavior-Driven-Development philosophy (BDD). In addition, three participants mentioned they perform ad-hoc testing. Finally, two of the participants combine the usage model with feedback from the end users.

Non-functional requirements were mentioned only in few cases as the drivers for designing test cases: robustness (2 answers), performance (1 answer), usability (1 answer), and different device configurations (1 answer). Other strategies mentioned by only one participant each are: code coverage, defining assertions in code, dependencies, testing the GUI model, and testing the business logic. 

We also found that developers tend to prefer a single criteria for designing test cases, as very few respondents reported more than one preference. Only 15 participants reported mixed strategies, as those described in the following responses:


\begin{myquote}\emph{
``l look for boundary conditions - i try to work out what happens in the Grey Areas - i look for ways of breaking it. i also test a range of use cases, how Can the user interact with the app? when combinations are possible? i try to test the most probable Scenarios and some strange ones.''
}\end{myquote}

\begin{myquote}\emph{
``1) specific tests depending on what the app should do. For instance: schedule should be as precise as is required for a scheduling app. 2) robustness: find the limits on the app by constantly changing the aspect ratio of the screen or switching app on and off. 3) look at CPU utilization while testing the app... app should not drain battery unnecessary''
}\end{myquote}

\begin{myquote}\emph{
``Based on the user stories or Use cases. Define initial state (local data). Perform scenario (call a rest service / perform an activity, etc). Validate final screen or rest service result''
}\end{myquote}

\begin{figure}[ht]
\begin{center}
\includegraphics[width=\linewidth]{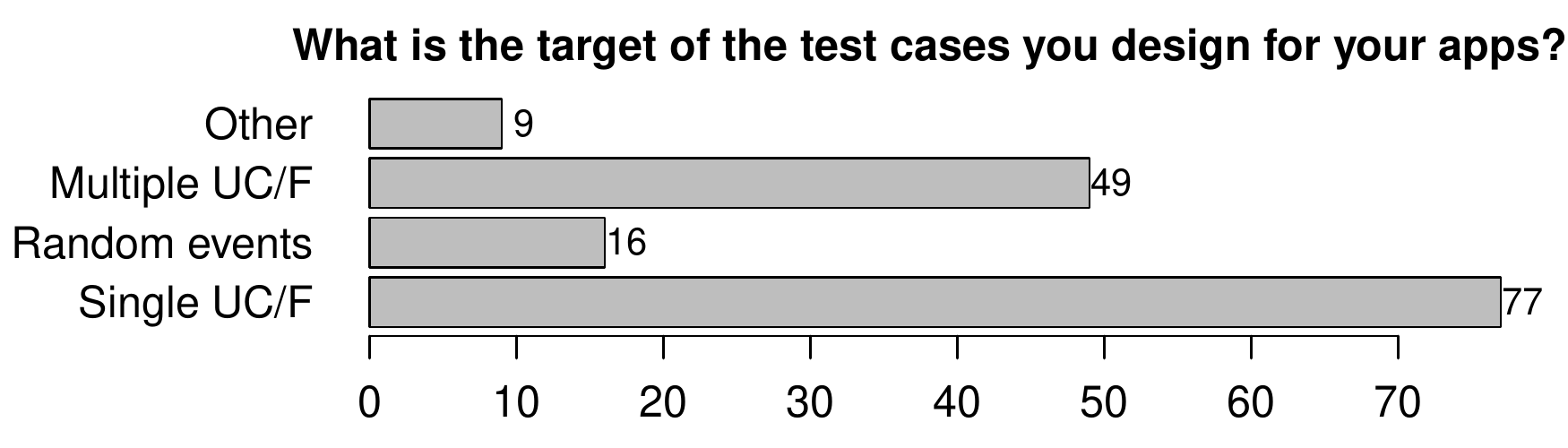}
\vspace{-0.9cm}
\caption{Target of test cases designed by the survey participants. The bar plots show the number of times each target was selected by the participants. *UC means Use Case, and F means Feature.}
\label{fig:test-cases-target}
\end{center}
\vspace{-0.3cm}
\end{figure}

The answers to \textbf{SQ4} were complemented by participants reporting the testing goals they have when designing test cases (\textbf{SQ5}). Fig.~\ref{fig:test-cases-target} depicts the responses for \textbf{SQ5}.  As shown in Fig.~\ref{fig:test-cases-target}, the developers prefer to design test cases that target individual uses cases/features (77 answers), or combinations of multiple uses cases/features (49 answers). Random events were mentioned only by 16 participants.  The option ``Other" was selected by nine participants; four of the ``Other'' answers were ``none'', two participants mentioned non-functional attributes (i.e., performance and robustness), one participant responded ``corner cases'', and one mentioned ``unit test''.
\vspace{-0.3cm}

\noindent \fbox{
  \parbox{0.94\linewidth}{
\textbf{Answer to RQ$_1$.} Mobile developers (as represented by our survey sample) heavily rely on usage models to document and design test cases. First, requirements are documented using different artifacts such as feature lists, informal natural language descriptions, user stories and use cases; only six out of 102 participants reported they do not use any artifact to document requirements. Second, the surveyed participants mostly rely on manual testing  and unit testing for their testing strategies. The rationale provided by the participants for their preference and effort dedication to manual testing is supported by several reasons such as (i) changing requirements, (ii) lack of time for testing and process decisions, (iii) size of the apps, (iv) lack of knowledge of the tools and techniques, (v) usability and learning curve of available tools, and (vi) the cost of maintaining automated testing artifacts.  Finally, the surveyed developers mostly focus on the usage model  to design test cases, and use one or a combination of use cases/features as the target for their test cases.}}

\subsection{RQ$_2$: What are MDs preferences for automatically generated test cases?}

\begin{figure}[t]
\begin{center}
\includegraphics[width=\linewidth]{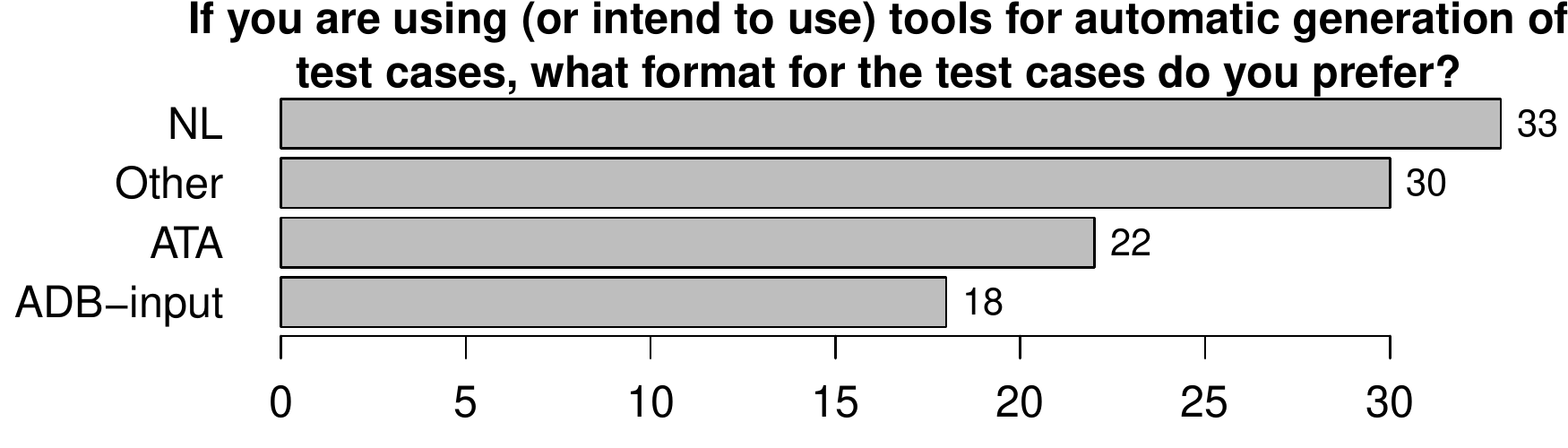}
\vspace{-0.9cm}
\caption{Automatically generated test cases format preferred by the survey participants. NL means Natural Language, ATA means Automated Test API, and ADB-input means input commands generated via the Android Debug Bridge (ADB).}
\label{fig:test-case-format}
\vspace{-0.8cm}
\end{center}
\end{figure}

Concerning \textbf{SQ6}, natural language  is the format preferred by the survey participants (See Figure~\ref{fig:test-case-format}) for automated test cases, with 33 out of 102 participants selecting this option. The second most popular answer was the ``Other'' option (30 participants), which represented mostly the lack of participant knowledge about tools for automatic generation of test cases or the lack of preference for any format. Unfortunately, the option ``Other'' did not provide us with actionable knowledge; 17 textual answers for the ``Other'' option are empty or claim no preference for any format; ten participants reported that they do not use/do not like/are not aware of tools for automatic generation of test cases; one answer was unclear; one answer mentioned scripts augmented with comments; and one participant responded ``Replayable event streams, starting from a known good state''.

\begin{figure}[t]
\begin{center}
\includegraphics[width=\linewidth]{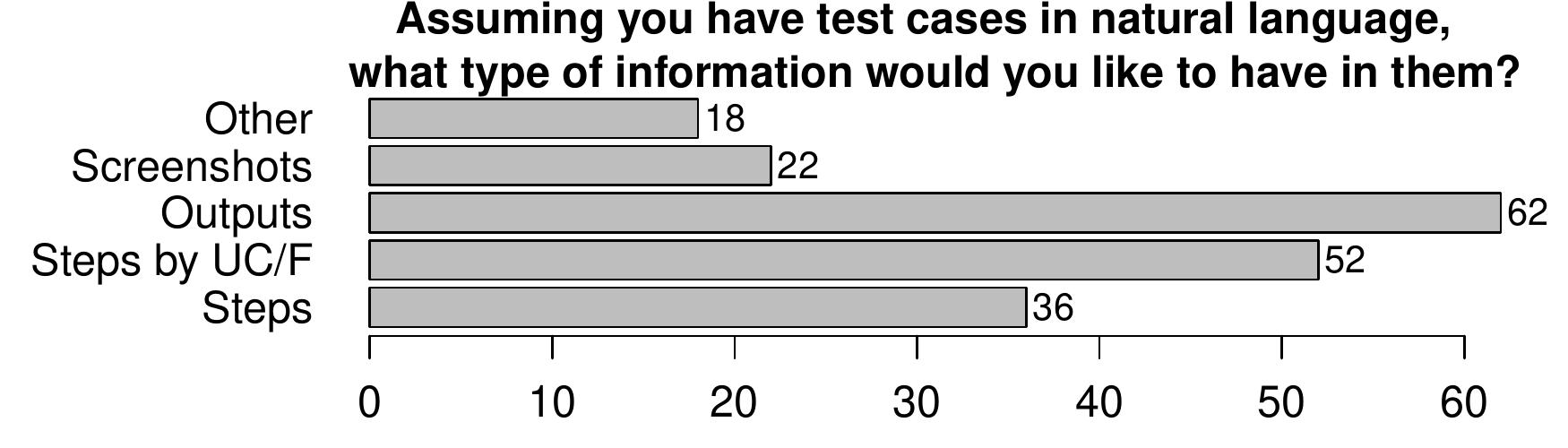}
\vspace{-0.9cm}
\caption{Information preferred by survey participants in automatically generated test cases. UC means Use Case and F means Feature.}
\label{fig:test-cases-info}
\end{center}
\vspace{-0.2cm}
\end{figure}

The third most popular option for \textbf{SQ6} was test cases written with automated testing APIs such as Espresso and Robotium (22 participants), and the least selected choice was ADB input commands (18 participants). Therefore by combining the number of participants voting for natural language test cases and test cases written with ATAs, the results suggest that Android developers prefer expressive test cases over low level scripts with input commands. 

Concerning the preferred information that developers would like to have in an ideal automatically generated test case (\textbf{SQ7}), expected outputs and reproduction steps organized by use case or feature are preferred over the other options. The answers for \textbf{SQ7} are depicted in Fig.~\ref{fig:test-cases-info}. 82 out of 102 participants (80.39\%) selected either ``Expected outputs'' or ``Reproduction steps grouped by use case/feature''. Only 21.7\% of the participants agreed on having screenshots as part of test cases. And, the textual answers to the ``Other"  option include:  reason/motivation for the test case, device and contextual information (e.g., Android OS version, display dimension, internet connection status), malicious user inputs, and specifications like in the RSpec framework for Ruby~\cite{RSpec}.

\noindent \fbox{
  \parbox{0.94\linewidth}{
\textbf{Answer to RQ$_2$.} Automatically generated test cases in natural language or expressed using automated testing APIs (e.g., Robotium or Espresso)  are preferred by the participants. This suggest a preference for high-level languages instead of low level events (e.g., using ADB input commands). In addition, the surveyed developers prefer to have test cases that include expected outputs and reproduction steps organized/grouped by use cases/features. 
}
}

\subsection{RQ$_3$: What tools are used by MDs for automated testing?}

\begin{figure}[t]
\begin{center}
\includegraphics[width=\linewidth]{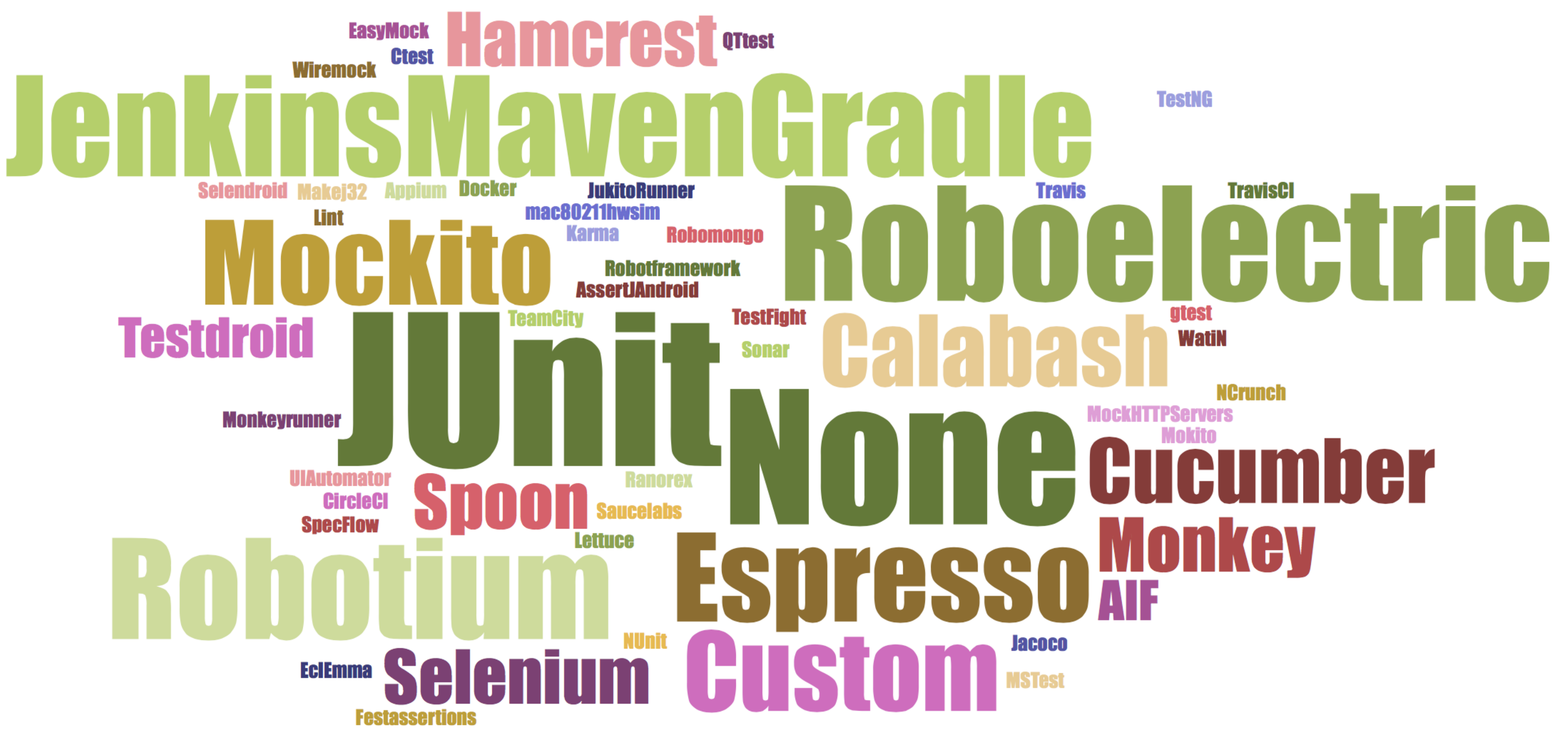}
\vspace{-0.9cm}
\caption{Tools used by the participants for automated testing of Android apps.}
\label{fig:tools}
\end{center}
\vspace{-0.1cm}
\end{figure}

\textbf{Tools for automated testing}. 55 different tools have been used by our survey participants (\textbf{SQ8}); the tools and frequencies are depicted as a word cloud in Fig.~\ref{fig:tools}. The most used tool is JUnit~\cite{JUnit} (45 participants), followed by Roboelectic~\cite{Roboelectric} with 16 answers, and Robotium~\cite{Robotium} with 11 answers.  28 participants explicitly mentioned they have not used any automated tool for testing mobile apps.  39 out of 55 tools were mentioned only by one participant each, which suggests that mobile developers do not use a well established set of tools for automated testing. In addition, surprisingly, Monkey~\cite{Monkey}, the state-of-the-art tool for fuzz/random testing, was mentioned only by three out of 102 participants, and the results are similar for other tools designed/promoted by Google: Espresso~\cite{Espresso}  (eight participants), MonkeyRunner~\cite{MonkeyRunner}(one participant), Lint~\cite{Lint} (one participant), and UIAutomator~\cite{Espresso} (one participant). None of the mentioned tools allow for automatic derivation of test cases from  source code, real app usages, or requirements specifications\footnote{It is worth nothing that the Android Monkey tool generates random sequences of events, however, the sequences are not easy to document or describe in terms of use cases.}
 
\textbf{Experiences with random testing tools}. 60.78\% (62 out of 102) of the participants reported no experience with Monkey or tools for random testing (\textbf{SQ9}). 15 participants (14.71\%) provided no-rationale/unclear answers. Concerning usefulness, 13 participants (12.75\%) consider Monkey and random testing as useful tools for corner cases or stress testing, and in some cases for finding performance issues. For instance:


\begin{myquote}\emph{
``Yes, it's useful to detect minor/stability issues. For instance it sometimes finds issues happened when tapping buttons many times quickly.''
}\end{myquote}

\begin{myquote}\emph{
``Monkey is very useful for stress testing the application or to verify that there are no leaks (typically memory leaks) that build up over time. Sometimes, it also catches the odd bug as well.''
}\end{myquote}


However, eight of those 15 participants pointed out that some issues or limitations are related to low impact of the discovered bugs and reproducibility:


\begin{myquote}\emph{
``They can be useful in finding memory leaks, ANRs, bad navigation flows and the like. They can be problematic in doing unexpected things, e.g. exiting your application during a test run. ''
}\end{myquote}

\begin{myquote}\emph{
``Good for stress testing, not very consistent results.''
}\end{myquote}

\begin{myquote}\emph{
``They add a lot of noise for very little signal. They are good at finding some really weird cases, but mostly it seems like the cases they find wouldn't normally be hit by a user and that time could be better spent adding features.''
}\end{myquote}

\begin{myquote}\emph{
``I used it long time ago and find that it works as a solution only for low-quality fragile code. It's rarely helps to improve overall quality of the app.''
}\end{myquote}


11 participants (10.78\%) provided answers in which the main message is that Monkey/Random testing tools are not useful because of reproducibility issues, maintenance costs of the scripts, and lack of tangible benefits:
\begin{myquote}\emph{
``hard to reproduce bugs, steps hardly reproducible by human beings''
}\end{myquote}
\begin{myquote}\emph{
``I've ran Android Monkey and it found some defects, but the developers said "that barely happens" or "that never happens". So the defects weren't looked into.''
}\end{myquote}
\begin{myquote}\emph{
``I don't think random tests provide any value. The ideal is to have a tool able to perform an entire session (run several scenarios)  / without having the test-runner to kill and initialize the app before running each scenario.''
}\end{myquote}


Although random testing has been proven to be ``useful" by the research community~\cite{Duran:TSE84,Arcuri:TSE12,Ciupa:2011,Amalfitano:Mobilesoft2015}, available random testing tools for mobile apps (e.g., Monkey) have issues such as lack of expressiveness. For instance, the Android Monkey tool  allows for reproducibility of event streams (by using the same seed), but it does not have log capabilities for creating a higher level representation of the streams. In addition, it seems to be the only tool available for the community for random testing of mobile apps. 

Researchers have also designed tools for mobile random testing (e.g., Dynodroid~\cite{Machiry:FSE13}), however those tools are still only known by the research community or have impediments for industrial usage (e.g., applicability only under certain conditions). In general, the main finding here is the lack of experience with and lack of usage of random testing tools by the surveyed participants. Regarding the usefulness, few participants reported cases in which the tools are useful for finding corner cases and stress testing; and few participants reported that the tools are not useful at all. However, common complaints in both cases (from both participants who saw benefits and those who didn't) are lack of reproducibility of the event sequences. These results suggest, that current tools used by Android developers need to be improved to allow expressiveness of the generated streams. Also more effort in the research community should be dedicated to promote the usage of random testing tools generated as part of research, and to deliver tools that can be easily adopted by the industry.

\vspace{0.2cm}
\noindent \fbox{
  \parbox{0.94\linewidth}{
\textbf{Answer to RQ$_3$.}
The surveyed participants rely on a diverse set of tools for supporting automated testing of mobile apps. In particular, 55 different tools were reported showing a preference for APIs such as JUnit, Roboelectric and Robotium. Compared to the study by Kochhar \emph{et al.}~\cite{Kochhar:ICST15}, we report a larger set of tools answered by a larger set of participants. However, both studies agree on listing JUnit, Roboelectric and Robotium as part of the top-4 used tools. Record \& replay, and random testing tools are used only by few participants. In the case of random testing tools, few participants claimed some benefits such as stress testing, execution/discovery of corner cases, and execution of events that are hard to generate by humans. However, impediments for increased adoption of random testing tools (e.g., the Android Monkey tool) are the lack of expressiveness of the generated event streams, and difficulty reproducing scenarios.
}
}

\subsection{RQ$_4$: Do MDs consider code coverage as a useful metric for evaluating test cases quality?} 14 out of 102 participants reported they do not use code coverage, do not use automation tools, or were unaware of code-coverage as a quality metric (\textbf{SQ10}); and  six  out of 102 participants provided no valid answers. From the remaining 82 participants, 51 answered ``No'' (i.e., code coverage is not useful),  29 answered ``Yes'' (i.e., code coverage is useful), and two participants provided a ``yes-no'' answer. In the case of the ``No'' answers, 19 out of 51 augmented the answer claiming that code coverage is not a good metric for measuring quality of test cases because there are other useful and better methods/metrics such as code (test cases) reviews, number of faults detected by the test cases (fault-detection capability), features covered by the test cases (feature coverage), or the ``works for me'' criteria\footnote{Note that the effectiveness of code coverage for measuring the quality of test suites has been already questioned by the software engineering community  \cite{HOLMES:ICSE2014,Just:FSE14,Zhang:FSE15}.}. Examples of the answers claiming that code coverage is not a useful metric for evaluating test case quality include the following:


\begin{myquote}\emph{
``No. We measure the number of uncaught bugs and regressions over time that devs had to spend time fixing''
}\end{myquote}

\begin{myquote}\emph{
``No, calculate total coverage based on features, covered elements etc.''
}\end{myquote}

\begin{myquote}\emph{
``I don't usually participate in the testing side of things, but I wouldn't use code coverage as a metric for quality as the two are completely distinct and different things.''
}\end{myquote}

\begin{myquote}\emph{
``Code coverage categorically does not measure the quality of tests. It is useful to show that code is not currently tested but it says nothing further about the code that is already under test. Many people -- probably most -- are quite skilled at writing useless tests.  Education is the only tool for producing high-quality tests and code review is the only tool for ensuring that quality. That said, fuzz testing, for instance, is a very powerful tool for certain kinds of testing. Knowing how to use the tools available to us is part of that education.''
}\end{myquote}


29 out of 102  participants found code coverage a useful metric for measuring test cases quality, in particular for identifying code entities that have not been tested. Examples of their answers are as in the following:


\begin{myquote}\emph{
``I use code coverage mainly as a tool to ensure I haven't forgotten any major areas of testing. Most of my projects have a minimum coverage requirement of 75-80\%''
}\end{myquote}

\begin{myquote}\emph{
``Yes, I try to keep code coverage at an acceptable level. This is definitely not the only thing that matters, but I think it does matter.''
}\end{myquote}

\begin{myquote}\emph{
``We use code coverage because it's easy to measure, it's a good enough metric, and because if developers feel they are being measured they are more likely to write more tests, thus generally producing the desired outcome.''
}\end{myquote}


Finally, the ``Yes-no'' answers claim code coverage is not useful at all, but they help to identify parts of the code that have not been tested: 


\begin{myquote}\emph{
``Yes and no. It's an indication if something is tested, not that the test is correct.''
}\end{myquote}

\begin{myquote}\emph{
``No. We use code coverage more as a guide to which part of the code base might need more attention in terms of writing more tests.  We don't really have other metric for measuring quality of the test cases.''
}\end{myquote}


\vspace{0.5cm}
\noindent \fbox{
  \parbox{0.94\linewidth}{
\textbf{Answer to RQ$_4$.} Code coverage is not used or not considered as useful for measuring the quality of test cases by 63.73\% (i.e., 65 out of 102) of the surveyed participants. Some of the reasons explaining the lack of confidence in the metric is that they prefer fault-detection or feature-coverage capabilities of test cases as a measure of quality, or they prefer to measure test cases quality by performing code reviews. On the other side, some participants consider that code measure is a useful metric because it helps to identify parts of the code that are not tested. 
}
}

%% file: threats.tex
Threats to \textsl{construct validity} concern the relationship between theory and observation, and relate to possible measurement imprecision when extracting data used in a study. To minimize this threat we filtered out incomplete surveys, participants with zero years of experience, and surveys with invalid answers. Moreover, to minimize a source of inexactitude in our study on the open questions, we followed a grounded theory-based approach \cite{CorbinStrauss}. In particular three of the authors performed an open coding by independently creating categories, then the codes were standardized and in case of no-agreement the answers were marked as ``Unclear".

Threats to \textsl{external validity }concern the generalizability of our
findings. The results in our study may not be generalizable to developers on other platforms, moreover our study only focuses on developers from open source projects on Github and we can not guarantee that all participants are commercial developers (although the average industrial experience for participants is between 5-10 years). In addition, the testing practices and automated tools used by our sample set may not generalize across all mobile developers. However, despite this fact, we believe the information provided by our respondent pool can be used to effectively provide guidelines for the research community towards devising more practical automated testing approaches. 

Another threat related to generalizability is that the results of our study are based on 102 respondents which might not be representative of the global community of Android developers. However, this study surveys a comparable number of Android developers to other studies \cite{Mona:ESEM13,Kochhar:ICST15,Aho:ICSTW14}.

%% file: related.tex
	In this section we present the related work and we differentiate the outcomes of our study compared to other studies concerned with investigating the topic of mobile testing.

	Erfani \emph{et al.} \cite{Mona:ESEM13} performed a study to understand the challenges that developers face during the life cycle of mobile software development. The study comprises interviews and a semi-structured survey targeted to 12 experts on mobile development and 188 people from the general mobile community respectively. Therefore, the findings can be categorized in four main topics: (i) general challenges such as fragmentation, testing support, open/closed platforms, data intensive, and frequent code changes; (ii) development across multiple platforms with problems such as native vs. hybrid apps, capabilities of platforms, code reuse vs. writing from scratch, behavioral consistency cross platform, and effort on migration across platform; (iii) current testing practices like manual testing, developers as testers, platform specific testing, levels of testing, beta testers; and finally (iv) testing challenges including limited unit test support for mobile specific features, better monitoring support, crash reports, emulators, missing platform-supported tools, rapid changes, multiple scenarios to validate, app stores and usability testing. 

	The study concluded that one of the most important challenges for developers is having to deal with multiple platforms, since the knowledge of one platform typically can't be transferred to another. In addition, tools to monitor and measure the performance of mobile apps are important for developers as are testing frameworks and tools. Our paper differs from the goals of this study in that we focus on examining the testing practices and preferences of open source developers whereas Erfani \emph{et al.} analyzed a variety of aspects of the entire software development process.

	Kochhar \emph{et al.} \cite{Kochhar:ICST15} conducted an empirical investigation into open source apps and two different surveys, the first one comprising three questions asked to 83 android developers, and the second comprising five questions as an improved version of the first study, posed to 127 windows app developers at Microsoft. The study includes questions to investigate techniques used to test apps, frameworks used, types of testing used, reasons for using testing tools, and challenges encountered during testing process. The authors concluded that Android apps are not properly tested since around 86\% of the apps do not contain any test cases. In addition, existing automated tools are not able to reach certain parts of code in mobile apps and are typically prohibitively difficult to use. Finally, the study found that developers are not aware of many existing testing tools. Our paper differentiates itself in the fact that we attempted to distill developer's testing preferences and practices for both manual and automated practices in order to inform the development of more practical automated tools, and our participants were Android developers.

	Choudhary \emph{et al.} \cite{Shauvik:2015} presented a comparison between test input generation techniques for Android applications. Choudhary et al. studied these tools applied to 60 real-world applications considering four different criteria: (i) ease of use, (ii) android framework compatibility, (iii) code coverage achieved, and (iv) fault detection. The authors concluded that random testing (specifically Android Monkey) surpasses all other automated techniques. In contrast to this study we surveyed developers to investigate trends on usage of automated testing tools and experiences with random testing tools.

	Linares-V\'asquez \emph{et al.} \cite{Linares-Vasquez:ICSME17}, recently conducted a survey of current tools, frameworks, and services available to support mobile testing practices. This survey draws comparisons between different testing techniques and solutions, describing the benefits, and delineating drawbacks and trade-offs between different approaches/tools.  Additionally, the work offers a forward-thinking vision for effective mobile testing along three principles: Continuous, Evolutionary, and Large-Scale.  While this work offers a valuable perspective on the current state and potential future of mobile app testing, it does not survey developers to understand current mobile testing trends.

	Aho \emph{et al.} \cite{Aho:ICSTW14}, presented an industrial evaluation of the \textit{Murphy tool} that models the graphical user interface to support several testing tasks during the software development cycle. The experiences presented in the paper were based on the evaluation of three software systems and three test engineers from industry. The Murphy tool decreased the time and effort of generating test cases from the model. The authors concluded that Murphy helped to minimize the tedious and repetitive work while creating manual test cases that involves analysis and verification from the tester. Compared to this study, we do not focus on the evaluation of one particular approach rather we surveyed open source developers about the usage of different automated testing tools, and preferences for ideal automated testing techniques.

%% file: concl.tex
In this paper we presented the results of an empirical study with 102 contributors of open source mobile apps hosted at GitHub. In particular, the study was conducted with a survey aimed at gathering information about their practices in-the-wild and preferences for (i) designing and generating test cases,  (ii) using automated approaches, and (iii) assessing the quality of test suites.

Our survey reveals highly relevant opinions of open source developers such as they rely primarily on usage models (e.g, use cases, user stories) of their applications when designing test cases, and they prefer high-level expressive automatically generated test cases organized around use-cases.  As of today, little effort has been devoted to include usage models \cite{Tonella:ICSE14,Linares:MSR15,Kowalczyk:ASEW15} during automated test cases generation for mobile apps; thus, usage models and expressive test cases should be considered as an important goal for automated approaches/tools for mobile testing. The survey results support the need for multi-models in model-based testing as suggested by Linares-V\'asquez \emph{et al.} \cite{Linares-Vasquez:ICSME17}. 

Another result  we would like to highlight is the fact that code coverage is not perceived by the survey participants as an important measure of test cases quality. While code coverage has been widely used by researchers to validate automated approaches for testing mobile apps, this result could be used as an insight that reinforce the discussion regarding code coverage utilty \cite{HOLMES:ICSE2014,Just:FSE14}. Additionally, this should spur the discussion and creation of new evaluation models --- for new testing approaches/tools --- that consider other criteria such as relevant fault detection capability (e.g. faults along heavily traversed parts of the app) and feature coverage. 

Finally, our survey confirms the fact that despite the plethora of tools proposed by the research community, the  state-of-the-practice for automated testing are automation APIs; manually written test cases with automation APIs are very fragile to changes in the GUI of the app under test test \cite{Mona:ESEM13,Linares-Vasquez:ICSME17}; even the official Google tool for random/fuzz testing (i.e.,  Monkey) has a low usage rate. In order to aid in technology transfer, researchers should consider developer preferences and workflows when designing and evaluating their approaches.  Such preferences can be gleaned from the developer responses in this paper, and include among others: (i) A need for automatically generated test cases to co-evolve with apps and features, (ii) low-overhead tools that tightly integrate into current (agile) development workflows, and (iii) expressive test cases that allow for easier debugging and traceability between test cases and features.  By taking such preferences into consideration, researchers should be able to design approaches that make a meaningful impact during real mobile testing practices.